\documentclass{aip-cp}

\usepackage[numbers]{natbib}
\usepackage{rotating}
\usepackage{graphicx}

\begin{document}

\title{Excited Baryon Structure Using Exclusive Reactions with CLAS12}


\author{Daniel S. Carman \\
{\normalsize \it (for the CLAS Collaboration)}\\
{\normalsize \it Jefferson Laboratory, 12000 Jefferson Ave., Newport News, VA 23606}\\
{\normalsize carman@jlab.org}}

\maketitle

\begin{abstract}
Studying excited nucleon structure through exclusive electroproduction reactions is an important 
avenue for exploring the nature of the non-perturbative strong interaction. Electrocouplings for 
$N^*$ states in the mass range below 1.8~GeV have been determined from analyses of CLAS $\pi N$, 
$\eta N$, and $\pi \pi N$ data. This work made it clear that consistency of independent analyses 
of exclusive channels with different couplings and non-resonant backgrounds but the same $N^*$ 
electro-excitation amplitudes, is essential to have confidence in the extracted results. In terms 
of hadronic coupling, many high-lying $N^*$ states preferentially decay through the $\pi \pi N$ 
channel instead of $\pi N$. Data from the $KY$ channels will therefore be critical to provide an 
independent analysis to compare the extracted electrocouplings for the high-lying $N^*$ states 
against those determined from the $\pi N$ and $\pi \pi N$ channels. A program to study excited 
$N^*$ decays to non-strange and strange exclusive final states using CLAS12 will measure differential 
cross sections to be used as input to extract the $\gamma_vNN^*$ transition form factors for the most 
prominent $N^*$ states in the range of invariant energy $W$ up 3~GeV in the virtually unexplored 
domain of momentum transfers $Q^2$ up to 12~GeV$^2$.
\end{abstract}

\section{INTRODUCTION}

Detailed spectroscopic studies of the nucleon excitation spectrum and the structure of these excited 
states have played a central role in the development of our understanding of the dynamics of the 
strong interaction. The concept of quarks that emerged through such studies led to the development of 
the constituent quark model~\cite{isgur,capstick} (CQM) in the 1980s. As a result of intense experimental 
and theoretical effort over the past 30 years, it is now apparent that the structure of the nucleon and 
its spectrum of excited states ($N^*$) are much more complex than what can be described in terms of models 
based on constituent quarks alone. At the typical energy and distance scales found within the $N^*$ 
states, the quark-gluon coupling is large. Therefore, we are confronted with the fact that quark-gluon 
confinement, and hence the dynamics of the $N^*$ spectrum, cannot be understood through application of 
perturbative Quantum Chromodynamics (QCD) techniques. The need to understand QCD in this non-perturbative 
domain is a fundamental issue in nuclear physics that the study of $N^*$ structure can help to address. 
Such studies, in fact, represent the necessary first steps toward understanding how QCD generates mass, 
i.e. how mesons, baryons, and atomic nuclei are formed.

Studies of low-lying baryon states, as revealed by electromagnetic probes at low four-momentum transfer 
($Q^2 < 5$~GeV$^2$), have revealed $N^*$ structure as a complex interplay between the internal core of
three dressed quarks and an external meson-baryon cloud. $N^*$ states of different quantum numbers have
notably different relative contributions from these two components, demonstrating distinctly different
manifestations of the non-perturbative strong interaction in their generation. The relative contribution
of the quark core increases with $Q^2$ in a gradual transition to a dominance of quark degrees of freedom
for $Q^2 > 5$~GeV$^2$. This kinematics area still remains almost unexplored in exclusive reactions. 
Studies of the $Q^2$ evolution of $N^*$ structure from low to high $Q^2$ offer access to the strong 
interaction between dressed quarks in the non-perturbative regime that is responsible for $N^*$ 
formation.

Electroproduction reactions provide for a probe of the inner structure of the contributing $N^*$ 
resonances through the extraction of the amplitudes for the transition between the incident virtual 
photon-nucleon state and the final $N^*$ state, i.e. the $\gamma_vNN^*$ electrocoupling amplitudes, which 
describe the physics. Among these amplitudes are $A_{1/2}(Q^2)$ and $A_{3/2}(Q^2)$, which describe the 
$N^*$ resonance electroexcitation for the two different helicity configurations of a transverse 
photon and the nucleon, as well as $S_{1/2}(Q^2)$, which describes the $N^*$ resonance electroexcitation 
by longitudinal photons of zero helicity. Detailed comparisons of the theoretical predictions for these 
amplitudes with their experimental measurements is the basis of progress toward understanding 
non-perturbative QCD. The extraction of the $\gamma_vNN^*$ electrocouplings is needed in order to gain 
access to the dynamical momentum-dependent mass and structure of the dressed quark in the non-perturbative 
domain where the quark-gluon coupling is large~\cite{vm1}. This is critical in exploring the nature of 
quark-gluon confinement and dynamical chiral symmetry breaking (DCSB) in baryons. 

Current theoretical approaches to understand $N^*$ structure fall into two broad categories. In the first 
category are those that enable direct connection to the QCD Lagrangian, such as Lattice QCD (LQCD) and 
QCD applications of the Dyson-Schwinger equations (DSE). In the second category are those that use models 
inspired by or derived from our knowledge of QCD, such as quark-hadron duality, light-front holographic 
QCD (AdS/QCD), light-cone sum rules (LCSR), and CQMs. See Ref.~\cite{review} for an overview of these 
different approaches. It is important to realize that even those approaches that attempt to solve QCD 
directly can only do so approximately, and these approximations ultimately represent limitations that 
need careful consideration. As such, it is imperative that whenever possible the results of these intensive 
and challenging calculations be compared directly to the data from electroproduction experiments over a 
broad range of $Q^2$ for $N^*$ states with different quantum numbers. Comparisons of the experimental 
results on the $\gamma_vNN^*$ electrocouplings to the theoretical predictions provide for crucial insights 
into many aspects of the dynamics, including confinement and DCSB, through mapping of the dressed quark 
mass function~\cite{vm2} and extractions of the quark distribution amplitudes for the $N^*$ states of 
different quantum numbers~\cite{vm3}.

\boldmath
\section{CLAS $N^*$ PROGRAM}
\unboldmath

The $N^*$ program is one of the key cornerstones of the physics program in Hall~B at Jefferson Laboratory
(JLab). The large acceptance spectrometer CLAS~\cite{mecking} was designed to measure photo- and 
electroproduction cross sections and spin observables over a broad kinematic range for a host of different 
exclusive reaction channels. Consistent determination of $N^*$ properties from different exclusive channels 
with different couplings and non-resonant backgrounds offers model-independent support for the findings.

To date photoproduction data sets from CLAS and elsewhere have been used extensively to constrain
coupled-channel fits and advanced single-channel models. However, data at $Q^2$=0 allows us to identify 
new states and determine their quantum numbers, but they tell us very little about the structure of these 
states. It is the $Q^2$ dependence of the $\gamma_vNN^*$ electrocouplings that reveals these details.
In addition, electrocoupling data is promising for both spectrum and structure studies as the ratio of 
resonant to non-resonant amplitudes increases with increasing $Q^2$. Finally, the electroproduction data 
are an effective tool to confirm the existence of new $N^*$ states as the data must be described by 
$Q^2$-independent resonance masses and hadronic decay widths.

The program goal is the study of the spectrum of $N^*$ states and their associated structure over
a broad range of distance scales through studies of the $Q^2$ dependence of the $\gamma_vNN^*$
electrocouplings. For each final state this goal is realized employing two distinct phases. The first 
phase consists of the measurements of the experimental observables, cross sections and spin observables, 
in as fine a binning in the relevant kinematic variables ($Q^2$, $W$, $\cos \theta_m^*$) as possible 
with the data. The second phase consists of developing advanced reaction models that fully describe the 
data in order to then extract the electrocoupling amplitudes for the dominant contributing $N^*$ states. 
Electrocoupling amplitudes for most $N^*$ states below 1.8~GeV have been extracted for the first time from 
analysis of CLAS data in the exclusive $\pi^+ n$ and $\pi^0 p$ channels for $Q^2$ up to 5~GeV$^2$, in 
$\eta p$ for $Q^2$ up to 4~GeV$^2$, and for $\pi^+ \pi^- p$ for $Q^2$ up to 1.5~GeV$^2$.

\begin{figure}[htbp]
\centerline{\includegraphics[width=300pt]{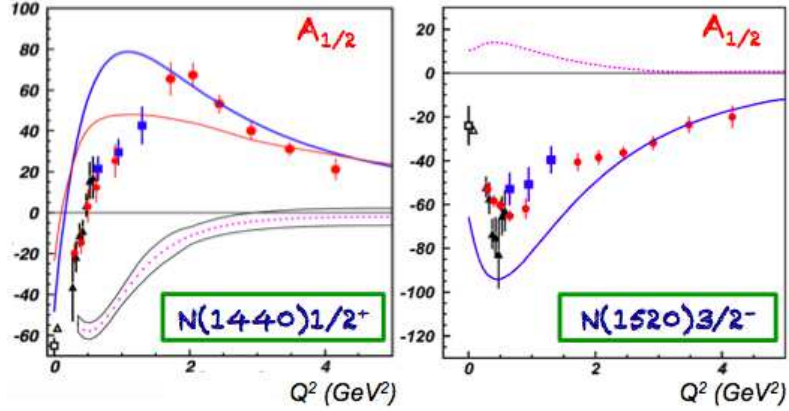}}
\caption{The $A_{1/2}$ electrocoupling amplitudes of the $N(1440)\frac{1}{2}^+$ (left) and 
$N(1520)\frac{3}{2}^-$ (right) $N^*$ states from the analyses of the CLAS $\pi N$ (circles) and 
$\pi\pi N$ (triangles, squares) data. (Left) Calculation from a non-relativistic light-front quark model 
with a running quark mass (red line) and calculation of the quark core from the DSE approach (blue line). 
(Right) Calculation from the hypercentral constituent quark model (blue line). The magnitude of the 
meson-baryon cloud contributions are shown by the magenta line in both plots. See Refs.~\cite{review,mokeev15} 
for details on the data and the models. The electrocouplings have units of 10$^{-3}$~GeV$^{-1/2}$.}
\label{low-lying} 
\end{figure}

Figure~\ref{low-lying} shows representative CLAS data for the $A_{1/2}$ electrocouplings for the
$N(1440)\frac{1}{2}^+$ and $N(1520)\frac{3}{2}^-$~\cite{review,mokeev15}. Studies of the 
electrocouplings for $N^*$ states of different quantum numbers at lower $Q^2$ have revealed a very 
different interplay between the inner quark core and the meson-baryon cloud as a function of $Q^2$. 
Structure studies of the low-lying $N^*$ states, e.g. $\Delta(1232)\frac{3}{2}^+$, $N(1440)\frac{1}{2}^+$, 
$N(1520)\frac{3}{2}^-$, and $N(1535)\frac{1}{2}^-$, have made significant progress in recent years 
due to the agreement of results from independent analyses of the CLAS $\pi N$ and $\pi\pi N$ final states
\cite{mokeev13}. The good agreement of the extracted electrocouplings from both the $\pi N$ and $\pi \pi N$ 
exclusive channels is non-trivial in that these channels have very different mechanisms for the non-resonant 
backgrounds. The agreement thus provides compelling evidence for the reliability of the results.

The size of the meson-baryon dressing amplitudes are maximal for $Q^2 < 1$~GeV$^2$ (see Fig.~\ref{low-lying}). 
For higher $Q^2$, there is a gradual transition to the domain where the quark degrees of freedom just begin 
to dominate, as seen by the improved description of the $N^*$ electrocouplings obtained within the DSE 
approach, which accounts only for the quark core contributions. For $Q^2 > 5$~GeV$^2$, the quark degrees of 
freedom are expected to fully dominate the $N^*$ states~\cite{review}. Therefore, in the $\gamma_vNN^*$ 
electrocoupling studies for $Q^2 > 5$~GeV$^2$ expected with the CLAS12 program, the quark degrees of freedom 
will be probed more directly with only small contributions from the meson-baryon cloud.

\begin{figure}[htbp]
\centerline{\includegraphics[width=430pt]{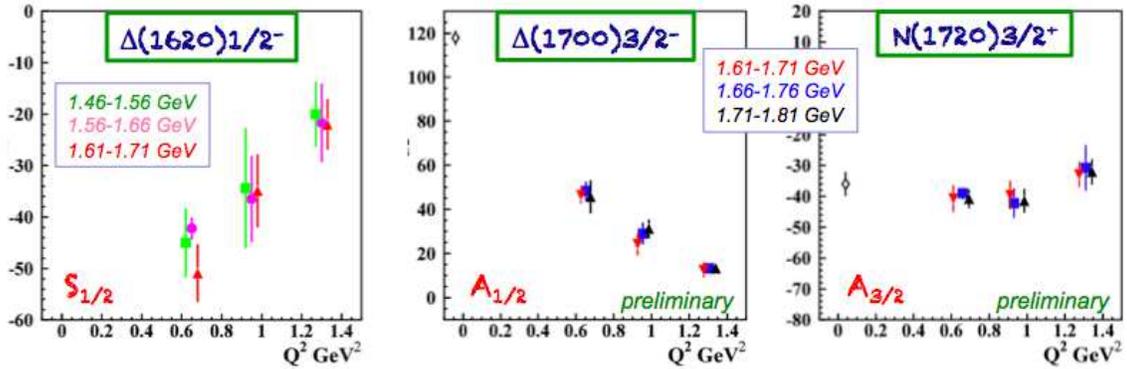}}
\caption{CLAS results of the electrocoupling amplitudes from analysis of the exclusive $\pi^+\pi^- p$ 
final state as a function of $Q^2$. (Left) $S_{1/2}$ of the $\Delta(1620)\frac{1}{2}^-$~\cite{mokeev15}, 
(middle) preliminary extraction of $A_{1/2}$ for the $\Delta(1700)\frac{3}{2}^-$~\cite{ect15}, and (right) 
preliminary extraction of $A_{3/2}$ for the $N(1720)\frac{3}{2}^+$~\cite{ect15} . Each electrocoupling 
amplitude was extracted in independent fits in different bins of $W$ across the resonance peak as shown 
for each $Q^2$ bin (points in each $Q^2$ bin offset for clarity). The electrocouplings have units of
10$^{-3}$~GeV$^{-1/2}$.}
\label{high-lying} 
\end{figure}

Analysis of CLAS data for the $\pi \pi N$ channel has provided the only detailed structural information 
regarding higher-lying $N^*$ states, e.g. $\Delta(1620)\frac{1}{2}^-$, $N(1650)\frac{1}{2}^-$, 
$N(1680)\frac{5}{2}^+$, $\Delta(1700)\frac{3}{2}^-$, and $N(1720)\frac{3}{2}^+$. Fig.~\ref{high-lying} 
shows a representative set of illustrative examples for $S_{1/2}$ for the $\Delta(1620)\frac{1}{2}^-$
\cite{mokeev15}, $A_{1/2}$ for the $\Delta(1700)\frac{3}{2}^-$~\cite{ect15}, and $A_{3/2}$ for the 
$N(1720)\frac{3}{2}^+$~\cite{ect15}. Here the analysis for each $N^*$ state was carried out independently 
in different bins in $W$ across the width of the state for $Q^2$ up to 1.5~GeV$^2$ with very good 
correspondence within each $Q^2$ bin. Note that most of the $N^*$ states with masses above 1.6~GeV decay 
preferentially through the $\pi \pi N$ channel instead of the $\pi N$ channel.

\begin{figure}[htbp]
\centerline{\includegraphics[width=330pt]{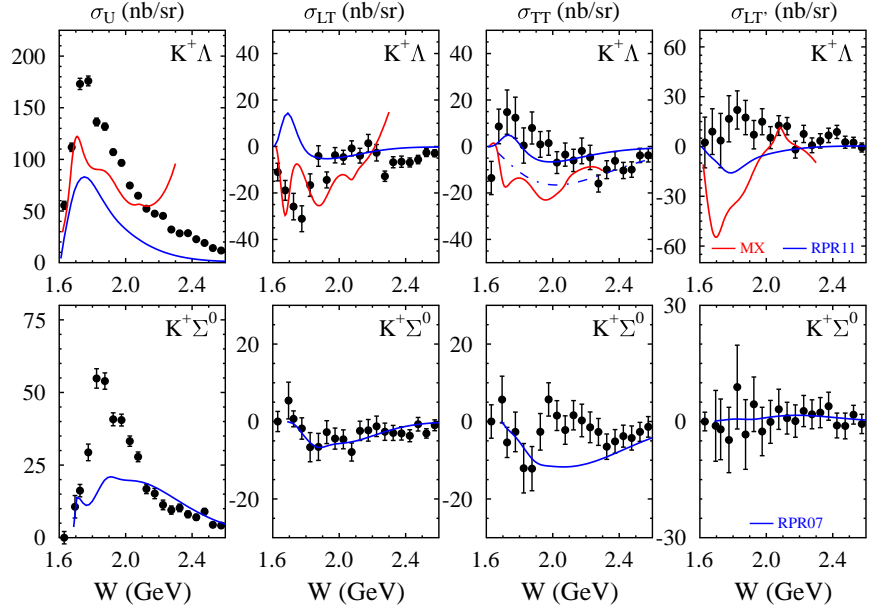}}
\caption{A small sample of the separated structure functions from CLAS data for the $K^+\Lambda$ (top)
and $K^+\Sigma^0$ (bottom) final states at $Q^2$=1.80~GeV$^2$ and $\cos \theta_K^*$=0.5 from Ref.~\cite{carman3}.
The curves show the isobar model from Maxwell~\cite{max12a} (red line) and the Regge plus Resonance model 
from Ghent~\cite{corthals} (for RPR-2007) and \cite{ryckebusch} (for RPR-2011) (blue lines) that were 
constrained by fits to the CLAS photoproduction data.}
\label{ky-data} 
\end{figure}

With a goal to have independent confirmation of the extracted electrocouplings for each $N^*$ state from
multiple exclusive final states, a natural avenue to investigate for the higher-lying $N^*$ states is the 
strangeness channels $K^+\Lambda$ and $K^+\Sigma^0$. In fact, data from the $KY$ channels is critical to 
provide an independent extraction of the electrocoupling amplitudes for the higher-lying $N^*$ states. The 
CLAS program has yielded by far the most extensive and precise measurements of $KY$ electroproduction data 
ever measured across the nucleon resonance region. These measurements have included the separated structure 
functions $\sigma_T$, $\sigma_L$, $\sigma_U = \sigma_T + \epsilon \sigma_L$, $\sigma_{LT}$, $\sigma_{TT}$, 
and $\sigma_{LT'}$ for $K^+\Lambda$ and $K^+\Sigma^0$~\cite{raue-car,5st,sltp,carman3}, recoil polarization 
for $K^+\Lambda$~\cite{ipol}, and beam-recoil transferred polarization for $K^+\Lambda$ and $K^+\Sigma^0$
\cite{carman1,carman2}. These measurements span $Q^2$ from 0.5 to 4.5~GeV$^2$, $W$ from 1.6 to 3.0~GeV, and 
the full center-of-mass angular range of the $K^+$. These final states, due to the creation of an $s\bar{s}$ 
quark pair in the intermediate state, are naturally sensitive to coupling to higher-lying $s$-channel resonance 
states at $W > 1.6$~GeV. Note also that although the two ground-state hyperons have the same valence quark 
structure ($uds$), they differ in isospin, such that intermediate $N^*$ resonances can decay strongly to 
$K^+\Lambda$ final states, but intermediate $\Delta^*$ states cannot. Because $K^+\Sigma^0$ final states can 
have contributions from both $N^*$ and $\Delta^*$ states, the hyperon final state selection constitutes an 
isospin filter. Shown in Fig.~\ref{ky-data} is a small sample of the available data in the form of the
$K^+\Lambda$ and $K^+\Sigma^0$ structure functions, illustrating its typical statistical precision.

Fig.~\ref{ky-data} includes two of the more advanced single channel models for the electromagnetic production
of $KY$ final states. The MX model is the isobar model from Maxwell~\cite{max12a}, and the RPR-2007
\cite{corthals} and RPR-2011~\cite{ryckebusch} models are the Regge plus Resonance framework developed
at Ghent. Both the MX and RPR models were developed based on fits to the extensive and precise photoproduction
data from CLAS and elsewhere and describe those data well. However, they utterly fail to describe the
electroproduction data in any of the kinematic phase space. Reliable information on $KY$ hadronic decays from 
$N^*$s is not yet available due to the lack of an adequate reaction model. However, after such a model is 
developed, the $N^*$ electrocoupling amplitudes for states that couple to $KY$ can be obtained from fits to 
the extensive existing CLAS $KY$ electroproduction data over the range $0.5 < Q^2 < 4$~GeV$^2$, which should 
be carried out independently in different bins of $Q^2$. The development of reaction models for the extraction 
of the $\gamma_vNN^*$ electrocouplings from the $KY$ electroproduction channels is urgently needed.

\boldmath
\section{CLAS12 $N^*$ PROGRAM}
\unboldmath
 
The electrocoupling parameters determined from the data involving the pionic channels for several 
low-lying $N^*$ states for photon virtualities up to $Q^2 \sim 5$~GeV$^2$ have already provided valuable 
information. At these distance scales, the resonance structure is determined by both meson-baryon 
dressing and dressed quark contributions. The $N^*$ program with the new CLAS12 spectrometer in 
Hall~B~\cite{clas12} is designed to study excited nucleon structure up to $Q^2$=12~GeV$^2$, the highest 
photon virtualities ever probed in exclusive electroproduction reactions. In the kinematic domain of $Q^2$ 
from 5 to 12~GeV$^2$, the data can probe more directly the inner quark core and map out the transition
from the confinement to the perturbative QCD domains.

The $N^*$ program with CLAS12 consists of two approved experiments. E12-09-003~\cite{e12-09-003} will 
focus on the non-strange final states (primarily $\pi N$, $\eta N$, $\pi \pi N$) and E12-06-108A
\cite{e12-06-108a} will focus on the strange final states (primarily $K^+\Lambda$ and $K^+\Sigma^0$). 
These experiments will allow for the determination of the $Q^2$ evolution of the electrocoupling parameters 
for $N^*$ states with masses in the range up to 3~GeV in the regime up to $Q^2$=12~GeV$^2$. These experiments
will be part of the first production physics running period with CLAS12 in 2017. The experiments will
collect data simultaneously using a longitudinally polarized 11~GeV electron beam on an unpolarized
liquid-hydrogen target at a nominal luminosity of $1 \times 10^{35}$cm$^{-2}$s$^{-1}$.

The program of $N^*$ studies with the CLAS12 detector has a number of important objectives. These include:

\vskip 0.3cm

i). To map out the quark structure of the dominant $N^*$ and $\Delta^*$ states from the acquired
electroproduction data through the exclusive final states including $\pi^0 p$, $\pi^+ n$, $\eta p$, 
$\pi^+\pi^- p$, $K^+\Lambda$, and $K^+\Sigma^0$. This objective is motivated by results from 
existing analyses such as those shown in Fig.~\ref{low-lying}, where it is seen that the meson-baryon 
dressing contribution to the $N^*$ structure decreases rapidly with increasing $Q^2$. The data can be 
described approximately in terms of dressed quarks already for $Q^2$ up to $\sim$3~GeV$^2$. It is therefore 
expected that the data at $Q^2 > 5$~GeV$^2$ can be used more directly to probe the quark substructure of 
the $N^*$ and $\Delta^*$ states~\cite{review}. The comparison of the extracted resonance electrocoupling 
parameters from this new higher $Q^2$ regime to the predictions from LQCD and DSE calculations will allow 
for a much improved understanding of how the internal dressed quark core emerges from QCD and how the 
dynamics of the strong interaction are responsible for the formation of the $N^*$ and $\Delta^*$ states 
of different quantum numbers.

\vskip 0.15cm

ii). To investigate the dynamics of dressed quark interactions and how they emerge from QCD. This work 
is motivated by recent advances in the DSE approach, which have provided links between the dressed quark 
propagator, the dressed quark scattering amplitudes, and the QCD Lagrangian. DSE analyses of the extracted 
$N^*$ electrocoupling parameters have the potential to allow for investigation of the origin of dressed 
quark confinement in baryons and the nature of DCSB, since both of these phenomena are rigorously 
incorporated into DSE approaches~\cite{review}.

\vskip 0.15cm

iii). To study the $Q^2$-dependence of the non-perturbative dynamics of QCD. This is motivated by studies of 
the momentum dependence of the dressed quark mass function of the quark propagator within LQCD~\cite{bowman} 
and DSE~\cite{bhagwat,roberts}. The calculated mass function approaches the current quark mass of a few MeV 
only in the high $Q^2$ regime of perturbative QCD. However, for decreasing momenta, the current quark acquires 
a constituent mass of 300~MeV as it is dressed by quarks and gluons. Verification of this momentum dependence 
would further advance understanding of non-perturbative dynamics. Efforts are currently underway to study the 
sensitivity of the proposed transition form factor measurements to different parameterizations of the momentum 
dependence of the quark mass~\cite{cloet}.

\vskip 0.15cm

iv). To access the quark distribution amplitudes in $N^*$ states of different quantum numbers based on the
LCSR approach and relating these amplitudes to the QCD Lagrangian within LQCD~\cite{vm3}.

\vskip 0.15cm

v). To offer constraints from resonance transition form factors for the $N \to N^*$ GPDs. We note that 
a key aspect of the CLAS12 measurement program is the characterization of exclusive reactions at high $Q^2$ 
in terms of GPDs. The elastic and $\gamma_vNN^*$ transition form factors represent the first moments of the 
GPDs~\cite{frankfurt,goeke}, and they provide for unique constraints on the structure of nucleons and 
their excited states. Thus the $N^*$ program at high $Q^2$ represents the initial step in a reliable 
parameterization of the transition $N \to N^*$ GPDs and is an important part of the larger overall CLAS12
program studying exclusive reactions.

\vskip 0.3cm

It is also important to note that the $\pi N$ and $\pi\pi N$ electroproduction channels represent the two 
dominating exclusive channels in the resonance region. The knowledge of the electroproduction mechanisms for 
these channels is critically important for $N^*$ studies in channels with smaller cross sections such as 
$K^+\Lambda$ and $K^+\Sigma^0$ production, as they can be significantly affected in leading order by 
coupled-channel effects produced by their hadronic interactions in the pionic channels. 

\section{CONCLUDING REMARKS}

The study of the spectrum and structure of the excited nucleon states represents one of the key physics 
foundations for the measurement program in Hall~B with the CLAS spectrometer. To date measurements
with CLAS have provided a dominant amount of precision data (cross sections and spin observables)
for a number of different exclusive final states for $Q^2$ from 0 to 4.5~GeV$^2$. From these data the
electrocouplings of most $N^*$ states up to $\sim$1.8~GeV have been extracted for the first time.

The $N^*$ program with the new CLAS12 spectrometer will extend these studies up to $Q^2$ of 12~GeV$^2$,
the highest photon virtualities ever probed in exclusive reactions. This program will ultimately focus
on the extraction of the $\gamma_vNN^*$ electrocouplings for the $s$-channel resonances that couple
strongly to the $\pi N$, $\eta N$, $\pi \pi N$, and $KY$ final states. These studies in concert with
theoretical developments will allow for insight into the strong interaction dynamics of dressed quarks
and their confinement in baryons over a broad $Q^2$ range. The data will address the most challenging
and open problems of the Standard Model on the nature of hadron mass, quark-gluon confinement, and
the emergence of the $N^*$ states of different quantum numbers from QCD.

\section{ACKNOWLEDGMENTS}

This work was supported by the U.S. Department of Energy. The author is grateful for many lengthy and
fruitful discussions on this topic with Victor Mokeev. The author also thanks the organizers of the
Hadron 2015 conference for the opportunity to present this work.



\end{document}